**TITLE PAGE**

**Title**: Performance of a deep learning system for detection of referable diabetic retinopathy in real clinical settings.

**Authors:** Verónica Sánchez-Gutiérrez[1], Paula Hernández-Martínez[1], Francisco J. Muñoz-Negrete[1,2], Jonne Engelberts[3], Allison M. Luger[3], Mark J.J.P. van Grinsven[3]

**Institutions:**
1 : Ophthalmology Department, Ramón y Cajal University Hospital, Ramón y Cajal Health Research Institute (IRYCIS), Madrid, Spain.
2: University of Alcalá de Henares, Madrid, Spain.
3: Thirona, Nijmegen, The Netherlands

**Corresponding author**: Mark J. J. P. van Grinsven
E-mail: markvangrinsven@thirona.eu

**ABSTRACT AND KEYWORDS**

**Background:** To determine the ability of a commercially available deep learning system, RetCAD v.1.3.1 (Thirona, Nijmegen, The Netherlands) for the automatic detection of referable diabetic retinopathy (DR) on a dataset of colour fundus images acquired during routine clinical practice in a tertiary hospital screening program, analyzing the reduction of workload that can be released incorporating this artificial intelligence-based technology.

**Methods:** Evaluation of the software was performed on a dataset of 7195 nonmydriatic fundus images from 6325 eyes of 3189 diabetic patients attending our screening program between February to December of 2019. The software generated a DR severity score for each colour fundus image which was combined into an eye-level score. This score was then compared with a reference standard as set by a human expert using receiver operating characteristic (ROC) curve analysis.

**Results:** The artificial intelligence (AI) software achieved an area under the ROC curve (AUC) value of 0.988 [0.981:0.993] for the detection of referable DR. At the proposed operating point, the sensitivity of the RetCAD software for DR is 90.53% and specificity is 97.13%. A workload reduction of 96% could be achieved at the cost of only 6 false negatives.

**Conclusions:** The AI software correctly identified the vast majority of referable DR cases, with a workload reduction of 96% of the cases that would need to be checked, while missing almost no true cases, so it may therefore be used as an instrument for triage.

**Keywords (3-10):** Automated detection, deep learning, diabetic retinopathy, screening, artificial intelligence

**List of abbreviations:**

DR: diabetic retinopathy

ROC: receiver operating characteristic

AUC: area under the curve

AI: artificial intelligence

**MAIN TEXT**

## 1. Introduction

Diabetes mellitus (DM) is a global health problem with a significant clinical impact on our society [1]. Sedentary lifestyles, obesity and lack of awareness are several potential factors that have contributed to an increased prevalence of DM, particularly in developing countries [2]. Diabetic Retinopathy (DR) is a common microvascular complication of DM and is a leading cause of acquired visual loss and blindness among the working-age population [3, 4, 5].

Early detection of this condition is critical to reach better outcomes, because DR may remain asymptomatic until it progresses to an advanced vision-threatening stage. For this reason, DR screening programs have long been recommended for patients with diabetes, adopting regular follow-ups, in order to detect the onset or progression of DR condition[6, 7]. Telemedicine-based screening for DR using digital nonmydriatic fundus photography has proven to be effective for detection of DR along with an appropriate referral [8, 9]. This method makes screening available to more patients and it can be used as a more efficient and cost-effective alternative than conventional in-office examination by an ophthalmologist [10].

As the prevalence of diabetes is expected to rise in the future, together with the aging of populations globally, an associated increase in DR cases should also be expected [11, 12]. In parallel, there is a need of intensify screening programs to include larger number of patients. So, in an effort to develop long-term strategies to manage this increasing burden of diabetes patients and possible retinopathy cases, artificial intelligence (AI) appears as an innovative tool of optimizing screening programs.

DR screening programs require human grader evaluation of the images which is resource-consuming, while deep learning approaches are gaining popularity in automated detection of DR from retinal fundus photographs as they have achieved excellent diagnostic performance in terms of high sensitivity and specificity [13, 14, 15].

In an attempt to increase the efficiency of DR screening and optimizing daily work flow, we designed this study to determine the performance of a deep learning system in detecting referable DR images from a real setting screening program compared with the evaluation done by ophthalmologist human graders. Our aim is to evaluate the amount of workload that could be released with this AI-based technology while keeping a suitable safety profile.

## 2. Materials and methods

### 2.1 Participants and images

A dataset of deidentified digital nonmydriatic fundus images were collected from consecutive recruited patients with a diagnosis of either type 1 or 2 DM who attended their regular visit at Ramon y Cajal Hospitals screening program. Fundus photographs were taken by trained nurses during routine clinical practice between February to December of 2019. Captures were done with a Topcon TRC-NW400 fundus camera (Topcon Medical Systems, Inc) using a 45º field of view. No mydriasis was applied. The acquisition protocol ensured that at least one fovea centered image per eye had to be taken. In total 7454 images of 3270 patients were collected. The mean age of the patients was 64.7 years (range 14-92 years) and 54% were male. The mean duration of their diabetes was 6.7 years (range 1-48 years) and 85% had type 2 DM.

All images were also scored for image gradeability, regarding contrast, clarity and focus, during the routine clinical practice and only gradable ones were included. Finally, 7195 images of 6325 eyes of 3189 patients were used in this study, see Table 1.

Retinal images were saved in a jpeg format and then forwarded through a safe telematic line to our tertiary care hospital. Images were anonymized prior to transfer and use in this study, following the ethical principles of the Declaration of Helsinki and the approval for the use of the deidentified images was obtained as all information collected was treated confidentially in strict compliance.

### 2.2 Automated grading for DR detection

RetCADv.1.3.1 (Thirona, The Netherlands) is a commercially available, Class IIa CE-marked medical device software that incorporates a deep learning framework that analyzes retinal images for the detection of DR related abnormalities in colour fundus images. The software is based on convolutional neural networks for the task of DR severity grading. In the process of analyzing the input colour fundus image, it compares regions in the image with regions extracted from normal and abnormal colour fundus images, which form the training data set of the software. None of the images included in the dataset for this study were used for training the system. The final outputs of the system are heat maps showing the locations of detected abnormalities (Fig.1) and a severity score in the range for DR. If the DR severity score is high, the case is deemed referable DR and the case should be referred for further testing. The DR severity score provided is a numerical index, varying from 0 to 100, where 0 represents the absence of retinopathy and an index closer to 100 indicates a high severity of DR is detected. The RetCAD system is calibrated in such a way that a case with a score >= 50 should be referred for further testing, while a case with DR severity score < 50 is not deemed referable. The performance of the RetCAD software has been directly compared with that of human experts in a separate validation study which was recently published [16].

## 2.3 Study protocol

To establish a reference standard (RS), the totally of the 6325 images went through a preliminary human grading performed during routine clinical practice by a trained ophthalmologist with over 5 years of experience reading fundus images (Grader 1). The grading for DR stage was based on the International Clinical Diabetic Retinopathy (ICDR) severity scale, with stages 1 (no DR), 2 (mild non-proliferative DR), 3 (moderate non-proliferative DR), 4 (severe non-proliferative DR) and 5 (proliferative DR) [17]. This RS was then adapted into referable and non-referable classification: images assigned with stages 1 or 2 were considered non-referable DR cases and those with stages 3, 4 and 5 as referable DR cases. Table 2 summarizes the distribution of DR disease severity in the study dataset regarding the corresponding reference standard. The prevalence of referrable DR was 1.5% in this study population.

RetCAD v1.3.1 was run on all of the images of the 6325 eyes and a DR score was reported. When multiple images were present for an eye, the maximum DR score of the multiple images was set as the DR score for that eye. A threshold point of 50for the software DR grading was used, so cases with a severity score of <50 were classified as non-referable DR cases and cases with a severity score >= 50 were classified as referable DR. All cases with a RetCAD score >= 50and cases were the software and Grader 1 did not agree were additionally annotated by a second independent human grader (Grader 2), with three years on DR screening experience.

## 2.4 Statistical analysis

Receiver operating characteristics (ROC) analysis was performed to measure the agreement between the RetCAD system and the reference standard. The RetCAD software was evaluated for the differentiation between non-referable and referable cases for DR. The ROC graph depicts the sensitivity and specificity pairs of the RetCAD software when using different cut-off threshold for the binary class problem. The overall performance of the RetCAD system is measured using the area under the ROC (AUC) value. Bootstrap analysis was used to calculate the 95% confidence interval for the ROC and AUC value. Sensitivity and specificity of the RetCAD software are also reported for the pre-defined cut-off threshold of 50.

## 3. Results

### 3.1 AI system outcome

Fig.2 shows a box plot of the DR severity scores obtained by the AI software and the ROC graph of the AI system for the detection of referable DR on eye-level. The RetCAD software obtained an AUC value of 0.988 [0.981:0.993] for the detection of referable DR. At the operating point with a cut-off of 50, the sensitivity of the RetCAD software for DR is sensitivity is 90.53 (86/95) eyes) and specificity is 97.13 (6051/6230).

An analysis of the disagreement between the reference grading and the RetCAD software, when using the pre-defined fixed threshold of 50 for DR for was made, where Grader 2 provided adjudication.

As can be seen in Fig.3, of the 6051 cases with a RetCAD score <50, only nine cases were deemed referable DR by Grader 1. Of these nine cases, three were graded as non-referable and six were graded as referable by Grader 2. All six cases were scored by RetCAD just below the threshold of 50 (scores: 41, 35, 42, 45, 42, 46). Five of these cases were marked as Mild DR by both graders whereas one case was scored as Severe DR.

Of the 274 cases scored as referable DR by the AI, Grader 1 indicated that 188 eyes were non-referable. Of these 188 cases, 38 were also graded as non-referable by Grader 2, whereas 150 were graded as referable by Grader 2. Of the 86 cases that were graded as referable by Grader 1, 81 of these cases were also graded as referable by Grader 2, whereas five cases were graded as non-referable. The 5 cases which were scored referable by the AI, referable by Grader 1, but non-referable by Grader 2, the AI scorings were (56, 57, 58, 54, 53), so just above the cut-off threshold of 50. All five cases were Moderate DR according to Grader 1, whereas two cases were No DR and three were Mild DR according to Grader 2.

With the given AI outcome, only 274 of the 6325 cases have to be checked for referable DR, whereas only six cases would have been missed. This is a workload reduction of 96%. Additionally in the flagged set of 274 cases, more referable DR cases were identified after adjudication of Grader 2.

## 4. Discussion

The results revealed that the RetCAD software achieved an AUC value of 0.988 (0.981-0.993) for the detection of referable DR cases based on the ICDR severity scale on a large dataset of nonmydriatic fundus images acquired during routine clinical practice. In our screening population, a cut-off threshold of 50 was selected resulting in an appropriate operating threshold, with values of sensitivity and specificity of 90.53 and 97.13, respectively. The software demonstrated a robust discrimination performance as we ensured that both performance indicators were above the screening guidelines recommendations [18, 19].

From our retrospective evaluation of RetCAD software, we also concluded that the system resulted in a large reduction of the workload for human graders. Concretely, only 4% of the images needed to be manually checked while missing 6 false negatives. DR screening programs can be optimized by reducing the number of fundus images requiring interpretation by human experts, who in return, can concentrate and spend more time reading abnormal images. An important advantage of RetCAD software is that the operating threshold can be calibrated to set an optimal cut-off point for different settings to meet particular operational requirements. A lower threshold can be selected to screen more conservative, i.e., higher detection rate at the cost of more false-positives, or a higher threshold which results in more workload reduction, at the cost of loss in sensitivity.This workload reduction of 96% displayed a huge reduction compared with previous studies [20, 21, 22, 23, 24] which have reported proportions ranged from 26.4% to 60%. However, the prevalence of referable DR was very low in this well controlled diabetic population, and this might have influenced on achieving such a high workload reduction. Even though in low DR burden settings, where the number of normal cases is typically much higher than the number of abnormal cases, the use of an automated system may result in increasing screening cost-effectiveness [25, 26].

It is important to test a DL system using independent datasets and in different populations, as this will assure the generalizability of the software in any clinical setting [19]. The RetCAD software has been evaluated on several datasets [16]. The data used in this study was collected on a large and highly representative sample in a real-world setting, as the patients were consecutively recruited in routine clinical practice from our DR screening program. Therefore, we believe that the automatic detection system analyzed in our study may be reasonably applied in other real clinical cohorts. However, integration into existing workflow remains challenging and prospective evaluation needs to be carried out to assess the discrimination performance of the system in normal procedure screening workflow.

Adoption of an AI tool in clinical practice requires a guarantee of its clinical utility. The development of automated detection routines can act as prefilters to flag out images with pathological lesions. An added value of RetCAD software, shown in our study, is that additional referral cases were identified, after Grader 2 scored them, when the software had already flagged these cases as first triage operator. We have to consider that AI systems can serve as a triage system and replace initial grading, but final verdict for treatment option will still be done by human specialists. Additionally, they may constitute a tool to strengthen screening programs. Automated detection systems offer an instant identification of patients with referable DR and allow increasing the number of people screened, which consequently, reduces delays in diagnosis of this treatable condition.

Legal liability in cases of misdiagnosis with AI is an issue that is yet to be resolved [27]. Consistent with this aspect, one of the main concerns of automated detection programs is missing referable cases with a potential delay in diagnosis and treatment. RetCAD missed six cases with referable DR according to both human graders; none of them had sight-threatening disease and all six cases were scored just below the threshold of 50 (Fig. 4). Of these six false negative cases, actually, the most worrisome was the one scored as severe DR by both Grader 1 and 2. Despite these false negative results, both human graders had a comparable or more number of missed cases.

Our study had some limitations. First, the preferred imaging protocol is based on seven stereo photographic fields [28], rather than the single macular field that was used in the present study. However, previous studies [29, 30, 31] have found this is an efficient DR screening method, as most DR changes usually occur in the posterior pole, although some may occur at the nasal retina and may not be able to be detected by the macular centered imaging protocol. Second, only the cases with disagreement between RetCAD and Grader 1, and with RetCAD DR score >= 50 were read by Grader 2. It could be that both RetCAD and Grader 1 have missed cases (in the 6042 cases set). This is considered in this study as the goal of the study was to see how AI can help reduce workload in the current DR settings. Any false negative in the 6042 cases would not have been detected in clinical practice as also Grader 1 would not have referred those cases. Finally, the low prevalence of referable DR cases in our study limits the comparison of our results to other populations with more disease prevalence.

In conclusion, RetCAD software was able to identify a large proportion of normal fundus images in a real DR screening setting at high sensitivity and could therefore be an instrument of triage. Using this system, attention can be drawn on potentially referral cases and this could result in a substantial workload reduction, without compromising patients safety. Future work should focus on integration and prospective evaluation of the software in screening workflow.

**DECLARATION**

This publication was supported by the Horizon 2020 EUROSTARS programme (Eurostars programme: IMAGE-R [#12712]). The funders had no role in the design and conduct of the study; collection, management, analysis, and interpretation of the data; and preparation and submission of the manuscript for publication. Mark van Grinsven, Jonne Engelberts, and Allison Luger are employees of Thirona, and Mark van Grinsven is also shareholder of Thirona. The dataset used and analysed during the current study are available from the corresponding author on reasonable request.

**FIGURE AND FIGURE LEGENDS**

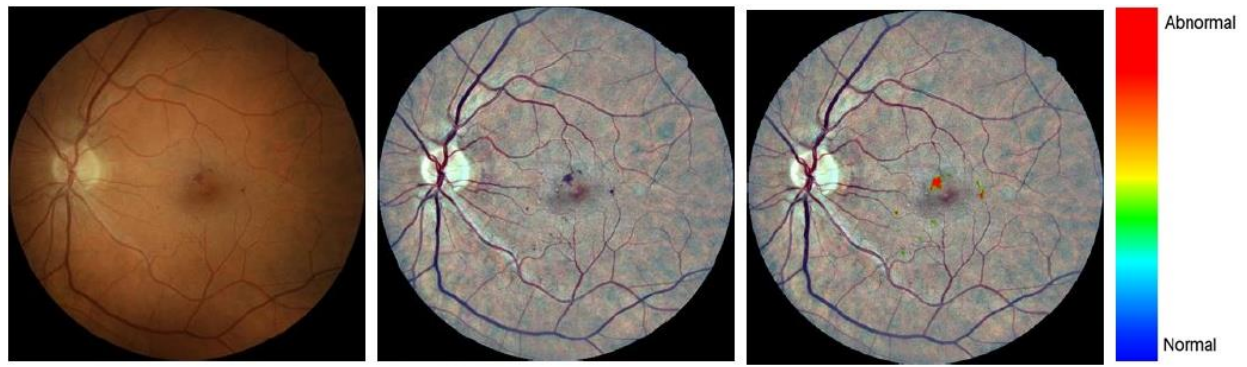

Fig.1. Left: original fundus image; middle: contrast enhanced image produced by RetCAD; right: DR heatmap produced by RetCAD.

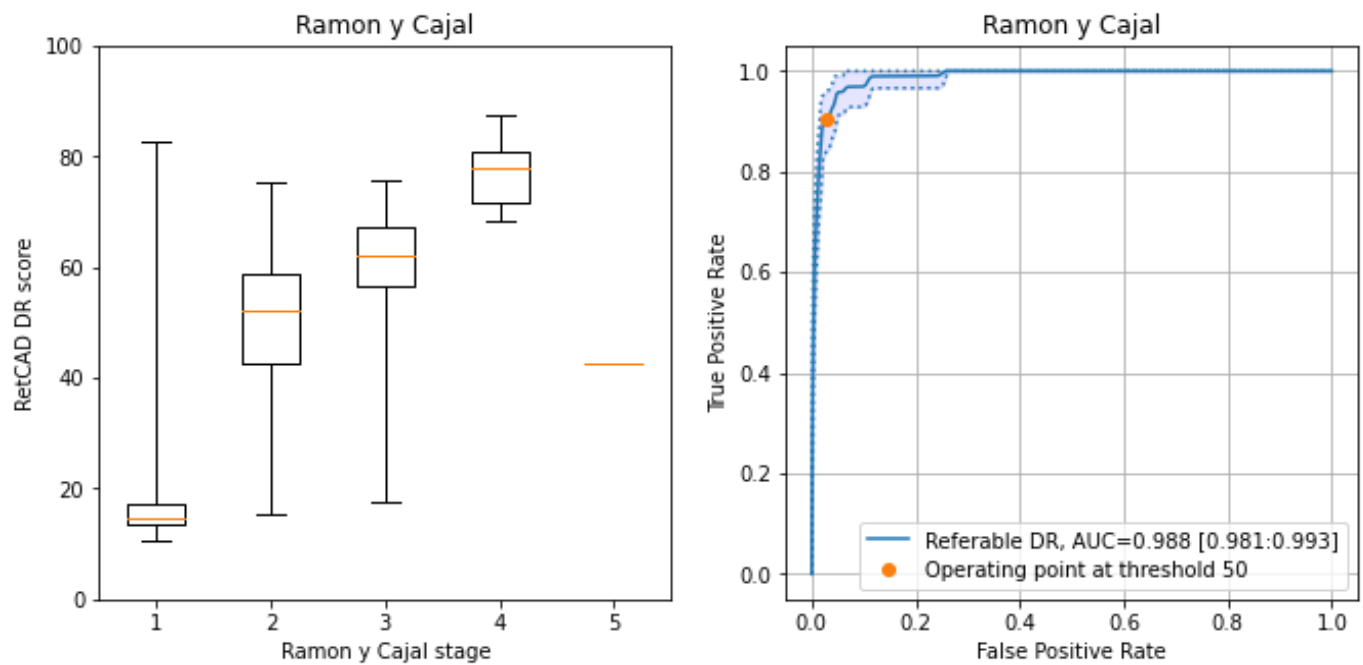

Fig.2. Left: Box plot of the RetCAD DR severity scores per DR severity category as set by the RS; Right: ROC of the RetCAD system for the detection of referable DR versus non-referable DR. The operating threshold with a cut-off of 50 is added in the plot.

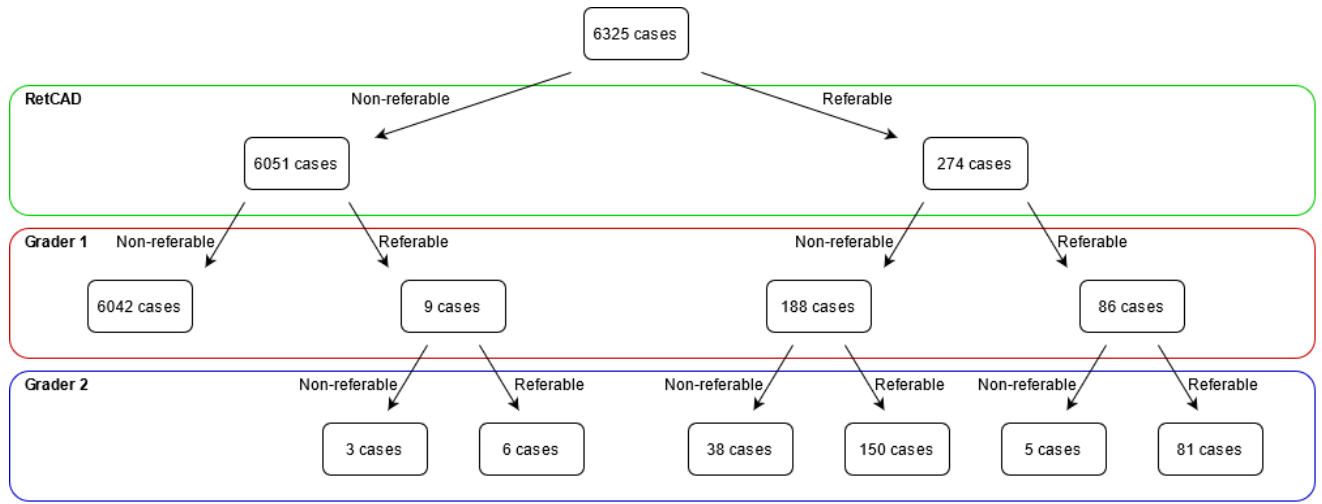

*Fig.3. Flowchart of case reading by the AI software, Grader 1 and Grader 2.*

**TABLES**

*Table 1. Distribution of sufficient image quality images.*

| Number of good quality images | 1 | 2 | 3 | 4 | 5 | 6 | total |
|---|---|---|---|---|---|---|---|
| Number of eyes | 5617 | 579 | 105 | 17 | 5 | 2 | 6325 |

*Table 2. Reference DR grading.*

| Diabetic retinopathy stage | 1 (no DR) | 2 (mild) | 3 (moderate) | 4 (severe) | 5 (proliferative) | Total |
|---|---|---|---|---|---|---|
| Number of eyes according to grader 1 | 6055 | 175 | 76 | 18 | 1 | 6325 |